\documentclass{rspublic}
\usepackage[colorlinks=true]{hyperref}
\usepackage{graphicx}
\usepackage{amsmath}
\usepackage{amssymb}
\usepackage{color}
\usepackage{bm}
\usepackage[round]{natbib}
\usepackage{amsfonts, amsthm, endnotes, setspace, verbatim, geometry, times, helvet, courier, url, dcolumn}
\usepackage{cleveref}

\usepackage{geometry}
\usepackage{subfigure}
\usepackage{amsmath,amssymb,amsthm}
\usepackage[mathscr]{eucal}
\usepackage[normalem]{ulem}

\usepackage{verbatim}

\def\half{{\textstyle \frac{1}{2}}}
\def\quarter{{\textstyle \frac{1}{4}}}

\usepackage{hyperref}
\usepackage{epsfig,graphicx}
\usepackage{dcolumn}
\usepackage{bm}

\begin{document}
\title[Understanding the Chain Fountain]{Understanding the Chain Fountain }

\author[J.S. Biggins and M. Warner]{J. S. Biggins$^{1,2}$, M. Warner$^{1,3}$}
 \affiliation{1.\ Cavendish Laboratory, 19 JJ Thomson Ave, Cambridge, UK, \\2.\ Trinity Hall, Trinity Ln, Cambridge UK, \\3.\ Rutherford School Physics Project, Cambridge, UK}

\date{\today}

\label{firstpage}		

\maketitle

\begin{abstract}{chain fountain bead siphon mechanics}
If a chain is initially at rest in a beaker at a height $h_1$ above the ground, and the end of the chain is pulled over the rim of the beaker and down towards the ground and then released, the chain will spontaneously ``flow'' out of the beaker under gravity. Furthermore, the beads do not simply drag over the edge of the beaker but form a fountain reaching a height $h_2$ above it. We show that the formation of a fountain requires that the beads come into motion not only by being pulled upwards by the part of the chain immediately above the pile, but also by being pushed upwards by an anomalous reaction force from the pile of stationary chain. We propose possible origins for this force, argue that its magnitude will be proportional to the square of the chain velocity, and predict and verify experimentally that $h_2\propto h_1$.
\end{abstract}

\maketitle

\begin{figure}\centering
\includegraphics[width=0.4 \textwidth]{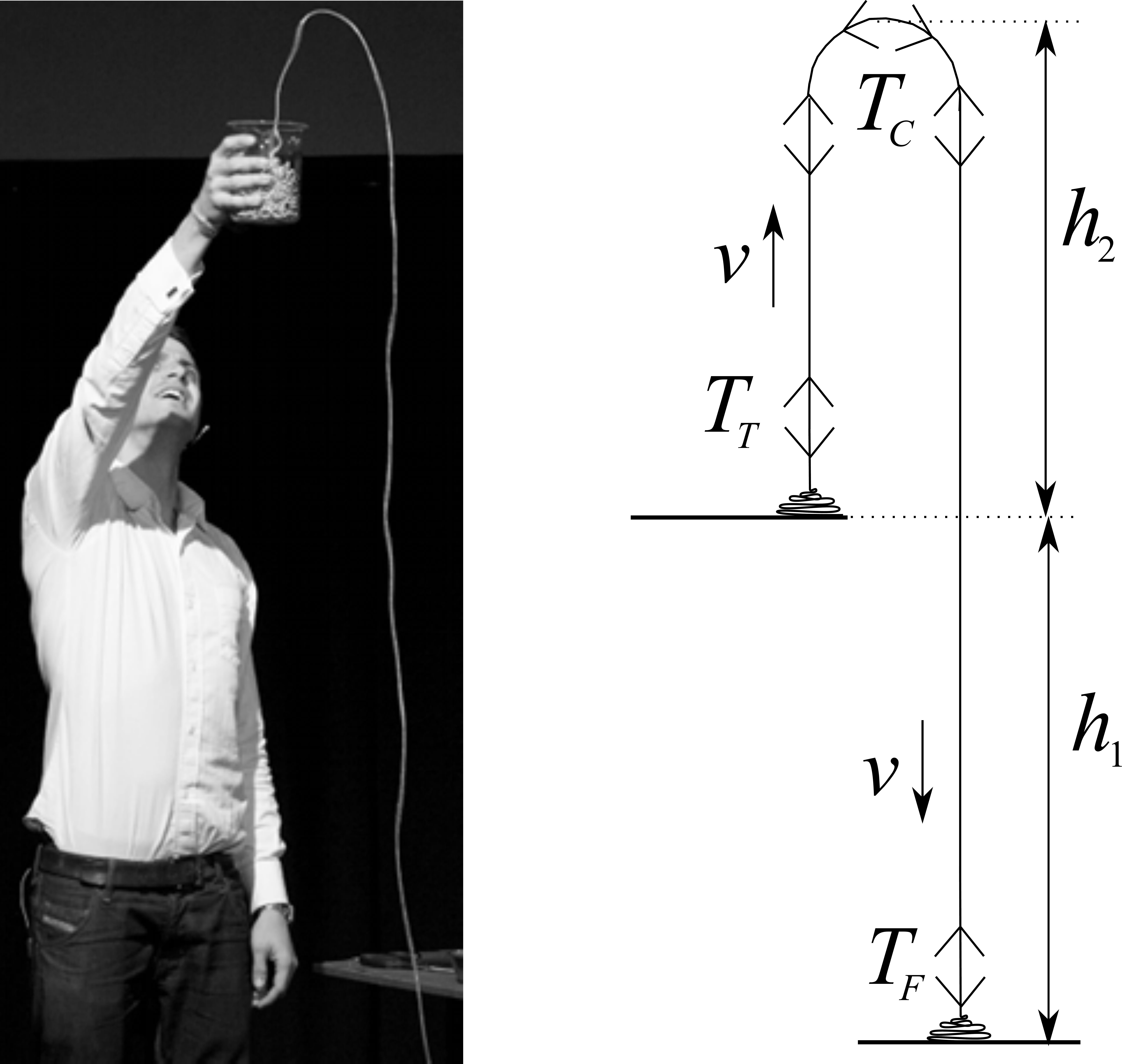}
\caption{Left: S.\ Mould demonstrating  a chain fountain. We thank J.\ Sanderson for permission to reproduce this photo. Right: Our minimal model of a chain fountain. A chain, with mass per unit length $\lambda$ is in a pile on a flat table a distance $h_1$ above the floor. It flows to the floor at a speed $v$ along the sketched trajectory, with $T_T$ being the tension just above the table, $T_C$ the tension in the small curved section at the top of the fountain and $T_F$ the tension just above the floor. }  \label{fig1}
\end{figure}

Chains are amongst the simplest, oldest and most ubiquitous of technologies. Since chains have great strength in tension but none in compression, their use typically requires a pile or spool of slack chain to be straightened into a tensile state by pulling on the end. One might imagine that this process would be comprehensively understood. However  \cite{mouldwebsite}  recently demonstrated that if a long chain is held in an elevated pot and the end of the chain is pulled down towards the ground, the chain will not only start to ``flow'' out of the pot and down to the ground, it will spontaneously leap above the rim of the pot forming a fountain, as shown in fig.\ \ref{fig1}. This is sufficiently surprising that several million people have viewed these videos. Here we show that these viewers are right to be surprised: to explain the existence of the fountain we must revisit traditional notions of how chains are picked up and conclude that the chain is not only pulled into motion by its own tension above the pile, but also pushed into motion by the pot.

The origin of the chain's flow from pot to floor is clearly the release of gravitational potential energy in a manner similar to a fluid siphon. The weight of the chain hanging between the pot and the floor pulls the chain out of the pot. The reason the beads leap above the pot remains unclear. We start our analysis with the simplest possible model of the chain fountain, sketched in fig.\ \ref{fig1}. Our model consists of a chain with mass per unit length $\lambda$, a pile of which is on a table at a height $h_1$ above the floor and is ``flowing'' down to the floor in a trajectory that is first vertically up for a height $h_2$, then reverses velocity in some small region of high curvature, then travels vertically down a distance $h_1+h_2$ before finally coming to rest in a second pile on the floor. In a steady state the moving part of the chain all moves at a speed $v$ along the chain's length. In our initial analysis, we assume the curved region at the apex of the fountain is small enough, and has tight enough curvature, that the centripetal acceleration is much larger than the gravitational acceleration ($v^2/r>>g$). The centripetal acceleration is then provided by the tension in the chain. If the (local) radius of curvature is $r$ and the tension in the curved region is $T_C$, then
\begin{equation}
T_C/r=\lambda v^2/r \quad \implies \quad T_C=\lambda v^2.\label{centrifugal}
\end{equation}
The cancellation of the radius of curvature from the above equation is quite remarkable. It indicates that a chain flowing along its own length can turn an arbitrarily sharp corner or trace any other shape provided the tension $T=\lambda v^2$. This eliminates the intuitive explanation for the chain fountain, namely that the chain must leave the beaker approximately vertically, and the inertia associated with this motion generates the fountain as the velocity cannot be immediately reversed. Given tension $T=\lambda v^2$ the chain could turn an arbitrarily sharp corner immediately above the rim of the pot. In reality the links prevent the chain from exceeding a large maximum curvature, but the curvature in the chain fountain is well below this limit. The cancelation of $r$ also means that a chain flowing along its own length can (in the absence of forces other than tension) form a steady state tracing any shape. This fact has been known since at least the 1850s: it features in Routh's dynamics textbook \citep{Routh_1860} and was known to the examiners of the 1854 Cambridge maths tripos \citep{Tripos_1854}.

The vertical portion of chain above the table is all moving at constant velocity so the forces on it must balance, giving
\begin{equation}
T_C=T_T+\lambda h_2 g,\label{abovetableforce}
\end{equation}
where $g$ is the acceleration due to gravity. The weight of the $h_2$ section of chain is $\lambda g h_2$, and $T_T$ is the tension just above the table. Similarly for the vertical portion above the floor we have
\begin{equation}
T_C=T_F+\lambda (h_2+h_1) g,\label{abovefloorforce}
\end{equation}
where $T_F$ is the tension just above the floor. In a time interval $\mathrm{d}t$ a mass of chain $\lambda v \mathrm{d}t$ is picked up from the table and acquires momentum  $\lambda v^2 \mathrm{d}t$. Traditionally we would expect this momentum to be provided by the tension $T_T$ in the chain immediately above the table, giving $T_T=\lambda v^2$. Combining this with eqns.\ (\ref{centrifugal}) and (\ref{abovetableforce}) we immediately see that $h_2=0$; there is no fountain. Similarly at the floor end of the chain in a time interval $\mathrm{d}t$ a mass of chain  $\lambda v \mathrm{d}t$ is brought to rest. Traditionally, since the chain is bought to rest by the floor, we expect this momentum to be provided by the floor, requiring that $T_F=0$. Then eqns~(\ref{abovetableforce}) and (\ref{abovefloorforce}) give $v=\sqrt{h_1 g}$.

The above treatment  of the chain fountain is highly dissipative. A unit length of chain releases potential energy $\lambda g h_1$ but only acquires a kinetic energy of $\half \lambda v^2=\half \lambda h_1 g$. This loss accords with the traditional view that picking up a chain with a constant force belongs to that class of dynamical problems where simple momentum conservation demands that half the work done is lost while half is converted into kinetic energy. This loss is implicit in Feynman's treatment of the problem \citep{feynman2013feynman} and discussed explicitly by the  \cite{RSPP}. Sand falling onto a moving conveyor belt is another example \citep{RSPP}, while charging a capacitor at constant voltage is an electrical analogue. One can see the dissipation in detail: If a chain with mass $\lambda$ per unit length is being picked up at a speed $v$ from a pile, the links are accelerated into motion by the tension $T_T$ just above the pile. In a time $\mathrm{dt}$ the length of chain picked up is $\mathrm{d}s=v \mathrm{d}t$, requiring a momentum $(\lambda v \mathrm{d}t)v=\lambda v^2 \mathrm{d}t$, so, $T_T=\lambda v^2$. The tension does work $T_T  \mathrm{d}s=\lambda v^2 \mathrm{d}s$ but the chain only receives kinetic energy $\half \lambda v^2  \mathrm{d} s$ so half the work is dissipated.

Since Mould did observe a fountain with a non-zero $h_2$ we must revisit these traditional ideas about chain pickup. For $h_2>0$ we must have $T_T<\lambda v^2$, so we must ask, if the tension in the chain is not providing all the momentum in the chain that is picked up, where is the rest coming from? The only possibility is that it is coming from the pile of chain and hence ultimately the pot. Accordingly we introduce an anomalous upwards force $R$ from the pile that acts on the part of the chain being brought into motion. Similarly we introduce an anomalous finite tension $T_F$ that acts to slow down the link being brought to rest on the floor. The momentum of the chain that is being picked up is now provided by both $R$ and $T_T$ giving
\begin{equation}
T_T+R=\lambda v^2.
\end{equation}

\begin{figure}\centering
\includegraphics[width=0.35 \textwidth]{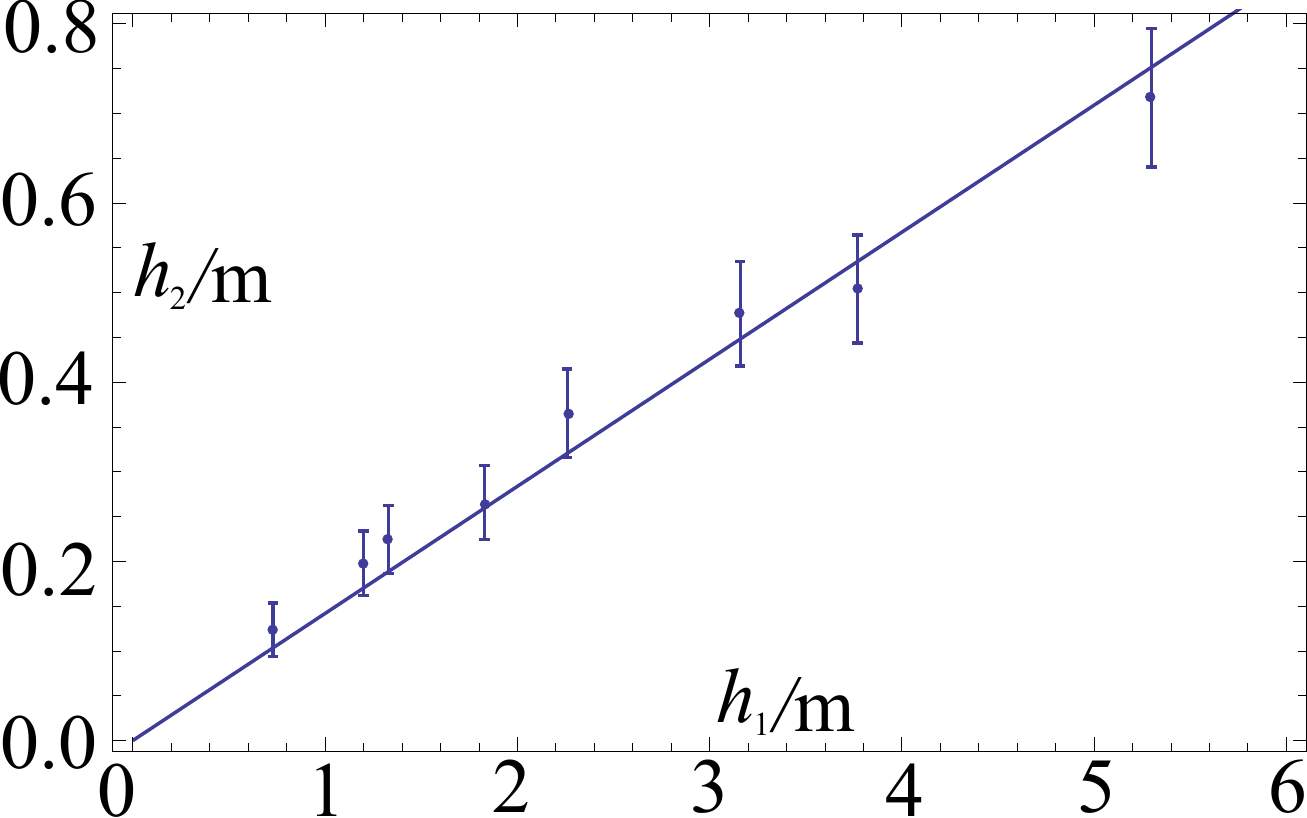}\caption{Fountain height $h_2$ plotted against drop $h_1$, the linear fit has $h_2=0.14 h_1$. The data was collected using  a 50m nickel-plated brass ball-chain consisting of balls of diameter 4.5mm connected by rods of length 2mm, shown in fig.\ \ref{fig2}. The chain was fed into a 1l plastic beaker in a random configuration, and the beaker was elevated to a distance $h_1$ above the ground. The end of the chain was  lowered from the beaker to the ground and then released, initiating the fountain. The fountain was filmed against a metre ruler to ascertain the fountain height $h_2$. Very small fountain heights were not measured because, if the fountain height is less than the height of the rim of the beaker, the fountain is supported by the beaker rather than free standing.}\label{expfig}
\end{figure}

On dimensional grounds all the forces must be proportional to $\lambda v^2$ so we write
\begin{equation}
R=\alpha \lambda v^2 \qquad T_F=\beta \lambda v^2 \label{definealphbeta},
\end{equation}
where setting $\alpha=\beta=0$ recovers the traditional view. We can now straightforwardly solve our equations to yield
\begin{equation}
h_2/h_1=\alpha/(1-\alpha-\beta), \qquad v^2=  h_1 g/(1-\alpha-\beta), \label{h2overh1}
\end{equation}
our principal results. One again sees that in the classical case $\alpha=0$ gives $h_2=0$, that is no fountain. Otherwise $h_2\propto h_1$. We verify this result experimentally in fig.\ \ref{expfig}.

A unit length of chain gains kinetic energy  $\half \lambda v^2 $ and releases potential energy $\lambda g h_1$. The ratio of the two is
\begin{equation}
KE/PE=\half v^2/(g h_1)=\frac{1}{2(1-\alpha -\beta)}\label{energyratio},
\end{equation}
Classically $\alpha=\beta=0$, this ratio is $1/2$ and half the potential energy is dissipated in the pickup process. Conservation of energy forbids the ratio in eqn (\ref{energyratio}) going above 1, requiring $\alpha+\beta\le\half$. The highest fountain would thus be made if $\alpha=\half$, $\beta=0$, yielding $h_2=h_1$ and no energy dissipation during the pickup process, however this bound is significantly in excess of the observed fountain heights of $h_2\sim 0.14 h_1$.

\begin{figure}\centering	
\includegraphics[width=0.6 \textwidth]{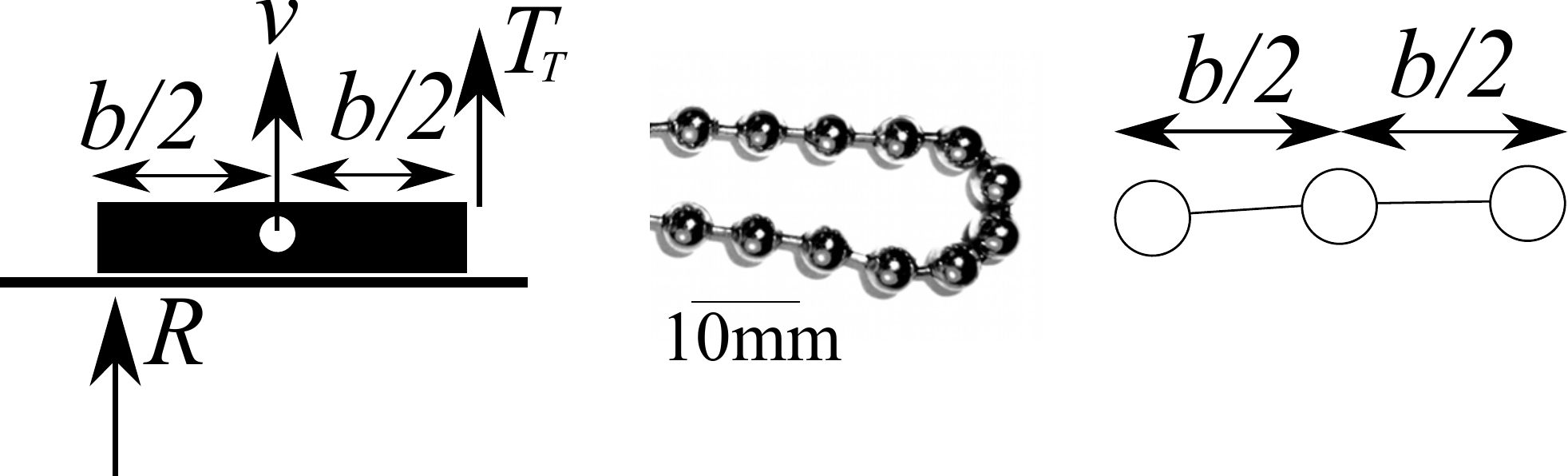}\caption{Left:A rigid rod of mass $m$ and moment of inertia $I$ lies on a horizontal surface (in practice the pile of chain) and is picked up via a vertical force $T_T$ applied at one end causing the rod's center of mass to rise at a speed $v$.  In order for the rod not to penetrate the surface, the surface must also provide a vertical reaction force $R$ on the opposite end of the rod. Middle and Right: The ball chain in our experiments required 6 beads to turn by $\pi$ (middle) so we model a link of the chain as consisting of 3 identical point masses (beads) connected by massless rods (right).}\label{fig2}
\end{figure}

We now consider the physical origin of the reaction force $R=\alpha \lambda v^2$ that gives rise to the chain fountain. At first sight it is highly counterintuitive that the origin of the fountain should be the pot pushing upwards on the chain. We model the chain as a set of freely jointed rods each with length $b$, mass $m$ and moment of inertia $I$. Links being set into motion are of necessity mostly at rest with a horizontal orientation, and are pulled at one end by preceding links that are directed largely upward. We model this via the situation in fig.\ \ref{fig2} where a single link is being picked up by a vertical force $T_T$ applied at one of its ends. This upward force  induces it to both rise and rotate. If the rod were in free space this motion would result in the other end of the link moving down. However if the rod is sat on a horizontal surface (modeling either the table or the rest of the pile of chain) then this cannot happen so the pile must supply an additional upwards reaction force $R$ acting at this end of the link.

In the initial stage there is a linear acceleration $a$ and an angular acceleration $\dot{\omega}$. To avoid the end at $R$ being rotated through the table, one needs $a=(b/2)\dot{\omega}$ and an $R$ acting to achieve consistent rates of change of linear and angular momentum. We use the initial force values as estimates of those acting. during pickup, thus
\begin{equation}
T_T+R=m a\quad \mathrm{and} \quad (T_T-R)(b/2)= I \dot{\omega},
\end{equation}
from which, (using $\dot{\omega}=a/(b/2)$)
\begin{equation}
R=\frac{1}{2} m a \left(1-\frac{I}{\quarter m b^2}\right).\label{fprime}
\end{equation}
Also, $T_T+R=m a$ must be the upward momentum flux $\lambda v^2$ on pick up. Replacing $m a$ in eqn (\ref{fprime}) we obtain
\begin{equation}
\alpha=\frac{R}{\lambda v^2}=\frac{1}{2}\left(1-\frac{I}{\quarter m b^2}\right).
\end{equation}
One can estimate $I$ of a link by recognizing that large bend in our chain is only achievable over about 6 beads, so we model one link in the chain as three connected beads (fig.\ \ref{fig2}).  The moment of inertia about the center of mass is $I=2 (m/3) (b/2)^2=m b^2/6$ which yields $\alpha=1/6$. If $\beta$ is small this yields $h_2\sim h_1/5$. Such rises are on the order of those observed.

The above estimate of $\alpha$ is somewhat crude, but in addition to the numerical estimate two general points emerge from it. Firstly, we have $R\propto \lambda v^2$, as we expect on dimensional grounds. Secondly we see that the precise value of $\alpha$ depends on the details of the chain: in the above it depends on the ratio $m b^2/I$ but in general it could depend on many details of the chain or indeed the conformation of the chain in the pile. We note that since $I\propto b^2$ for any link, $\alpha$ (and hence the fountain height) does not depend on the absolute size of the link ($b$) so our model would still apply in the continuum or string limit $b\to0$, and string-fountains have indeed been observed \citep{blundelfinkyoutube}. The above mechanism is only one way to generate a non-zero $\alpha$. An alternative would be to recall that our chain consists of spheres connected by rigid rods. When a segment of chain starts to move it may sometimes first move along its tangent (that is horizontally) resulting in the beads being dragged across the pile of stationary beads. This would result in collisions between the beads in the horizontally moving part and the stationary pile beneath that would kick the horizontally moving beads in the vertical direction, resulting in a non zero $R$.

It has previously been observed \citep{hamm2010weight, grewal2011chain} that if two equal lengths of hanging chain are released, one to fall in free space and one onto a table, the former accelerates at $g$ while, astonishingly, the latter can accelerate faster than $g$. Acceleration faster than $g$ can only be explained by an additional downward force acting on the chain, that is, a non zero $\beta$. The effect is the opposite of that discussed above and was very pronounced for the extreme type of chain in \cite{grewal2011chain}. Our chain is very different, (though \cite{hamm2010weight} consider a bead chain) and in a similar but crude experiment we did not observe this effect, suggesting that for our chain $\beta$ is small. More importantly, the functional form of eqn\ (\ref{h2overh1}) means that the fountain height does not much depend on $\beta$ --- it is $\alpha$ not $\beta$ that creates the fountain.

We tested our theory, that an anomalous reaction force from the pot accounts for the chain fountain, by conducting a simple numerical experiment. We modeled the chain using point masses connected by stiff springs (that did not otherwise interact). An ordered pile of chain was arranged on a horizontal ``table'' with which the masses collided elastically, and the end of the chain extended to the floor, which was modeled by a highly viscous liquid which brought the masses to a stop. The system was then integrated using a Verlet scheme and did indeed produce a chain fountain. Our numerical model did not have sufficiently stiff springs to accurately approximate an inextensible chain, so we cannot perform a quantitative comparison between numerics and theory, but we were able to test the main hypothesis of our paper, namely that for a fountain there must be an $\alpha>0$ producing an upward force from the table during the pickup process. In our numerics, elastic collisions between the masses and the ``table'' provided this force. We then removed the ``table'' and instead supported the chain numerically by having a region of zero gravity around the pile.  The chain still flowed to the ground, but there was no fountain. This is expected as in this case there was no possible mechanism to produce an $\alpha$, and confirms that to make a fountain the pot (or table) must push on the chain.

Some may worry that our model of the chain fountain is simply too crude to predict a fountain without an $\alpha$. A much more complete approach would be to solve for the full shape of the fountain. We imagine the fountain is in a steady state so that the moving part of the chain all moves at speed $v$ along the direction of the chain  and that, a distance $s$ along the chain, the chain makes an angle $\theta(s)$ with the vertical. Force balance parallel to the tangent of the chain then gives
\begin{equation}
T'(s)=\lambda g \cos{\theta} \label{paralelcatenary}
\end{equation}
while perpendicular to the tangent of the chain the forces must provide the necessary centripetal acceleration for the links in the chain going round the local curvature $1/r=\theta'(s)$ meaning
\begin{equation}
T(s) \theta'(s)+ g \lambda  \sin{\theta}=\lambda v^2 \theta'(s) .\label{centrifugalcatenary}
\end{equation}
If $v$ were zero, these equations would be the same as those for a chain hanging under gravity, and would be solved by it taking a catenary form. If the chain is moving, we see in the above equation that this simply results in a constant offset $\lambda v^2$ to $T(s)$ and that the equations are still solved by any inverted catenary \citep{airy1858mechanical}.  Thus we expect the chain to take on the form
\begin{equation}
y=d \cosh{\left((x-l)/d\right)}+c.
\end{equation}
We note this is an example of a more general result. If a chain hangs in a stationary equilibrium under the action of forces that only depend on position then, if the chain is induced to move along its length with a speed $v$, the tension will increase by $\lambda v^2$, and the original  shape of the chain will not be changed by the movement \citep{Tripos_1854}. The challenge with the chain fountain is to apply boundary conditions to fix these constants. Classically we would again set $T_T=\lambda v^2$ at the table end of the catenary while at the top of the catenary (where $\theta=\pi/2$), eqn\ (\ref{centrifugalcatenary}) gives $T_{C}=\lambda v^2-g \lambda /\theta'(s)\le \lambda v^2$. Also eqn.\ (\ref{paralelcatenary}) guarantees that $T_{C}\ge T_T= \lambda v^2$. The only way to reconcile these inequalities is to have the top of the catenary happen with $\theta'(s)\to \infty$ (that is infinitely sharply) and immediately above the table, in which limit both equations give $T_{C}=T_{T}=\lambda v^2$. Placing the top of the catenary immediately above the table corresponds to no fountain. It is possible to build $\alpha$ and $\beta$ into the catenary analysis. Curiously this is not enough to fix all the unknowns in the solution, one must also specify an additional boundary condition corresponding to the angle the chain leaves the pot at which, in practice, can be fixed by tilting the pot. In the limit of small angle (that is nearly vertical pickup) the catenary analysis reduces exactly to the simple model sketched in this paper.

Our treatment of the chain fountain is far from complete. The pick-up of the chain involves a point of rise that constantly traverses the chain pile. A transverse component to pickup results in links being picked up with some transverse velocity and requires the consideration of transverse forces. These rapidly changing transverse velocities lead to the formation of high amplitude transverse waves in the section of chain immediately above the pickup point. These waves contain some of the energy that is being dissipated during the pickup process, and their wavelength and amplitude are probably related to the conformation of the chain in the pot. The connection between the conformation of the chain, these waves and the observed value for $\alpha$ remains an interesting problem. Strikingly the waves appear almost stationary in video footage. This last property is easily explained: the wave speed on a chain is given by $\sqrt{T/\lambda}$ where $T$ is the tension. At the apex of the fountain we have $T_C=\lambda v^2$ and hence the wave speed and the actual speed of the chain are the same. Backward propagating waves will appear frozen at the apex. Between the pot and the apex $T$ is close to $\lambda v^2$ so the waves are almost stationary.

\cite{Santagelochainarch} have recently discussed a highly interesting problem: the formation of an arch when a long chain was arranged in ordered rows on a table and the end of the chain was then pulled at constant velocity along the table. Despite the apparent similarity to our fountain, there is an obvious difference: they pull perpendicular to the axis of the arch while we pull parallel. In their geometry the formation of any arch is a surprise (since the chain could remain in the plane of the table) which Hanna and Santangelo explained by showing that, in a region of rising tension between the (zero tension) rows and (high tension) moving portion of the chain, a perturbation transverse to the table will tend to steepen rather than flatten. They propose that this effect leads to the formation of structures perpendicular to the table which in turn lead to the arch because they are rectified by the table. The rectifying effect is analogous to our $R$. The arches they observe and analyze are sufficiently small that gravity is negligible compared to the centripetal acceleration and, since they do not reach a steady state, they do not analyze the dimensions of the arches. In contrast, in our chain fountain the beaker imposes the existence of a rising and a falling leg so rather than thinking about initiation we have focused on predicting the height of the steady state, for which gravity is critical. We suspect that, in their embryonic form, arch and the fountain are rather different structures. In the traditional view when a chain is picked up half the work done by the pick-up force is associated with motion of the chain transverse to the pickup direction. The fountain depends on reducing this effect via an anomalous reaction force \emph{parallel} to  the chain-pickup direction which augments the velocity of the chain in the pickup-direction and reduces transverse motion, whereas the arch requires a ``rectification''  force \emph{perpendicular} to the pickup-direction which organizes and augments the transverse motion. However, it is likely that if the chain-arch gets big enough its pickup direction will become effectively vertical and the two structures will converge to a common catenary. In any event, Hanna and Santangelo's work illustrates, as ours does, that the apparently simple process of pulling on the end of a pile of chain is far from understood and may yet yield further surprises.

Our central result --- that when a chain is picked up from a pile the picking-up force is augmented by a reaction from the pile --- may have consequences far beyond the chain fountain we study here. It will increase the rate at which any pile of chain is deployed when a force is applied at its end, and it will increase the energetic efficiency of the deployment process, which may have consequences for industrial design in areas as disparate as textiles and shipping. In areas such as space engineering (e.g. satellite tethers and space elevators) where efficiency is central, it may be worth maximizing the effect so that chain can be deployed with minimal expenditure of force and energy.

\section{Acknowledgements}
JSB thanks Trinity Hall, Walter Scott and the 1851 Royal Commission, MW's contribution is supported by the Rutherford School Physics Project of the Department for Education. We also thank Prof. J.\ D.\ Biggins, Miss A.\ Biggins and Mr J.\ Zamirski for helping with our experiments, Drs M. J. Rutter, D. Corbett and T. Tokieda for illuminating discussions and Dr T. Tokieda in particular for introducing us to the falling chain problem and anomalous forces. We are grateful to Profs. J. A. Hanna and C. D. Santangelo for telling us about the work of Routh, and for extensive discussions about their work.

\bibliographystyle{plainnat}

\end{document}